\documentclass[letterpaper,twoside]{article}
\begin{document}
\def\l{{\cal L}}
\def\h{{\cal H}}
\def\a{{\bigcirc\kern-0.86em \land\kern+.33em}}
\def\z{{\cal Z}}
\def\p{\perp}
\def\s{\Sigma}
\def\pro{{\rm P}}
\def\i{{\rm i}}
\def\e{{\rm e}}
\def\C{\hspace{.2em}{\kern-.1em{\raise.47ex\hbox{
$\scriptscriptstyle |$}}\kern-.40em{\rm C}}}
\def\be{\begin{equation}}
\def\ee{\end{equation}}
\def\ba{\begin{array}}
\def\ea{\end{array}}
\newcommand{\es}{\\[4mm]}
\newcommand{\lp}{\left(}
\newcommand{\rp}{\right)}
\newcommand{\lac}{\left\{}
\newcommand{\rac}{\right\}}
\newcommand{\lcr}{\left[}
\newcommand{\rcr}{\right]}
\newtheorem{theorem}{Theorem}
\newtheorem{definition}{Definition}
\newtheorem{lemma}{Lemma}
\newtheorem{proposition}{Proposition}
\newtheorem{question}{Question}
\newtheorem{example}{Example}
\newtheorem{corollary}{Corollary}
\newtheorem{remark}{Remark}
\def\demo#1{\vspace{-2.1855mm}\bigskip\noindent{\it Proof #1:}\nobreak\hskip0.5cm}
\def\cqfd{{\parskip=0pt\par\nobreak\rightline
{\vrule height 1.2ex width 1.4ex depth +.1ex}}}
\pagestyle{myheadings} \markboth{\hfill B. Ischi\hfill}{\hfill
PROPERTY LATTICES FOR INDEPENDENT QUANTUM SYSTEMS\hfill}

$ $

\vspace{2.6mm} \begin{center} {\bf PROPERTY LATTICES FOR
INDEPENDENT QUANTUM SYSTEMS}

\vspace{3.9mm}

{\large Boris Ischi} \vspace{1.2mm}

{\small Department of Mathematics and Statistics, McGill
University, }

{\small 805 Sherbrooke West, Montreal, Canada H3A 2K6}

{\small (e-mail: ischi@kalymnos.unige.ch)}

\vspace{2.7mm}
\end{center}
\vspace{4.1mm}

\begin{list}{}{\setlength{\leftmargin}{10mm}}
\item{\footnotesize We consider the description of two independent quantum systems by
a complete atomistic ortho-lattice (cao-lattice) $\l$. It is known
that since the two systems are independent, no Hilbert space
description is possible, {\it i.e.} $\l\ne\pro(\h)$, the lattice
of closed subspaces of a Hilbert space (theorem \ref{nohilbert}).
We impose five conditions on $\l$. Four of them are shown to be
physically necessary. The last one relates the orthogonality
between states in each system to the ortho-complementation of
$\l$. It can be justified if one assumes that the orthogonality
between states in the total system induces the
ortho-complementation of $\l$. We prove that if $\l$ satisfies
these five conditions, then $\l$ is the separated product proposed
by Aerts in 1982 to describe independent quantum systems (theorem
\ref{thetheorem}). Finally, we give strong arguments to exclude
the separated product and therefore our last condition. As a
consequence, we ask whether among the ca-lattices that satisfy our
first four basic necessary conditions, there exists an
ortho-complemented one different from the separated product.}

\item{\small{\bf Keywords:} quantum mechanics, independent systems, property lattices.}
\end{list}
\section{Motivations and notations}

In ordinary quantum mechanics, a system is described by a
(separable) Hilbert space over the complex numbers. The state
space is given by $\s=\h^*/\C$. Moreover, to any yes-no experiment
$\alpha$ on the system corresponds a $\mu(\alpha) \subset\s$ with
$\mu(\alpha)^{\p\p}=\mu(\alpha)$ (a closed subspace) such that the
answer ``yes'' is certain ({\it i.e.} the answer ``no'' is
impossible) if and only if the state of the system is in
$\mu(\alpha)$. Finally, the map $\mu$ is assumed to be surjective.

When two quantum systems are independent, Einstein, Podolsky and
Rosen point\-ed out that no Hilbert space description for the
total system is possible [5]. As a consequence the mathematical
description in the sense of Birkhoff and von Neumann [3] of that
situation appears as a natural question. To this end, we need a
generalization of the Hilbert space framework: Let $Q$ the set of
all possible yes-no experiments on a system $S$ at a certain time
$t$. Let $\s$ be a set (the state space) and $\l\subset 2^\s$ a
set of subsets of $\s$ such that there is a surjective map
$\mu:Q\rightarrow \l$ with the property that the answer ``yes''
for $\alpha$ is certain if and only if the state of $S$ is in
$\mu(\alpha)$. Then, following Aerts [1], we will assume that $\l$
is a p-lattice ($\l$ is called the property lattice)\newpage

\begin{definition}\rm Let $\s$ be a set and $\l\subset2^{\s}$. We call $\l$ a
p-lattice if\hfill\break (1) $\emptyset,\s\in\l$,\hfill\break (2)
$\cap a_\alpha\in\l$ for any family of elements of
$\l$,\hfill\break (3) $\{p\}\in\l$, $\forall p\in\s$.
\end{definition}

\begin{remark}\rm A p-lattice is a complete atomistic
lattice (say ca lattice). The set of atoms is given by $\{\{p\};\
p\in\s\}$. A complete atomistic ortho-lattice (say cao-lattice) is
ortho-isomorphic to the p-lattice $\{a\subset\s;\ a^{\p\p}=a\}$,
where $\s$ is the set of atoms.\end{remark}

\noindent (1) Define $I$ the trivial yes-no experiment by: ``Do
nothing on $S$ and answer yes'', and $O=I^\sim$, that is $I$ with
answers ``yes'' and ``no'' inverted. Then clearly
$\mu(O)=\emptyset$ and $\mu(I)=\s$. (2) Further, let $\alpha_i$ be
a family of yes-no experiments on $S$. Define $\pi\alpha_i$ by:
``choose freely an $\alpha_i$ and perform it''. Then
$\mu(\pi\alpha_i)=\cap\mu(\alpha_i)$. (3) Finally, for $p\in\s$
define $a_p:=\cap\{a\in\l;\ p\in a\}$. Then $p\in a_p$ and
$\varepsilon_p:=\mu^{-1}([a_p,\s])$ is the set of certain yes-no
experiments ({\it i.e.} the answer ``yes'' is certain) when the
state of the system is $p$. Suppose now that $\{p\}\ne a_p$. Let
$p\ne q\in a_p$. Then $\varepsilon_p\subset \varepsilon_q$. We
want to assume that when the state of the system changes, some
yes-no experiments become certain and some others do not remain
certain.

Finally, it is usually assumed that $\l$ has an
ortho-complementation. Note that there was an attempt to justify
this axiom [1], based on the following natural symmetric
anti-reflexive binary relation on $\s$:

\be p\p q\Leftrightarrow \exists\ \alpha\in Q;\ p\in\mu(\alpha)\
\mbox{and}\ q\in\mu(\alpha^\sim)\label{orthaerts}\ee where
$\alpha^\sim$ is the same yes-no experiment as $\alpha$ but with
switched  answers. It is in general delicate to give physical
arguments for this relation to induce an ortho-complementation on
$\l$, see [1].

\vspace{0.2cm} The time evolution of a system is given by a map
$u:\s_{t_0} \rightarrow\s_{t_1}$. W. Daniel [4] pointed out that
$u$ must satisfy

\be u^{-1}(b)\in\l_{t_0},\ \forall b\in\l_{t_1}\
,\label{daniel}\ee since $\mu(\alpha)=u^{-1}(\mu(\beta))$ for any
$\beta$ with $\mu(\beta)=b$, where $\alpha$ is the yes-no
experiment on the system at time $t_0$ defined by: ``evolve the
system from time $t_0$ to time $t_1$ and perform $\beta$''.

\begin{proposition}
Let $\l_1$, $\l_2$ be p-lattices and $f:\s_1\rightarrow\s_2$.
Assume that $f$ satisfies condition
(\ref{daniel}). Then $g(a):=\vee f(a)$ is $\vee$-preserving and equals $f$
on the atoms.
\end{proposition}

\demo{} Let $\{a_\alpha\in\l_1\}_{\alpha\in\omega}$, then since
$f$ satisfies condition (\ref{daniel}), we have $\vee a_\alpha
\subset$ $f^{-1}(\vee g(a_\alpha))$, that is $g(\vee
a_\alpha)\subset\vee g(a_\alpha)$. Moreover, since $g$ preserves
the order, $\vee g(a_\alpha)\subset g(\vee a_\alpha)$.\cqfd

\vspace{0.2cm} Let $S_1$ and $S_2$ be two physical systems
described by two p-lattices $\l_1$ and $\l_2$. Suppose that at a
given time $t_0$, the two systems are independent. This means
that\break\newpage\noindent at time $t_0$ any experiment on one
system does not alter the state of the other system (and, in
particular, that the two systems do not interact at time $t_0$).
It is the case in many experiments, for instance before the
interaction begins between two systems prepared in two independent
parts of the experimental device. Let $\l_{ind}$ a p-lattice
describing the physical properties of the total system $S$ at time
$t_0$ ({\it i.e.} $\l_{ind}$ is the property lattice of $S$). Then
we will assume that:

\begin{definition}\label{thedefinition}\rm
Let $\l_1$, $\l_2$ and $\l$ be p-lattices. Denote by ${\rm
Aut}(\l_i)$ the set of automorphisms of $\l_i$. We say that $\l$
is \hfill\break (P1) if $\s=\s_1\times\s_2$, \hfill\break (P2) if
$a_1\times\s_2\cup\s_1\times a_2\in\l$, $\forall a_i\in\l_i$,
\hfill\break (P3) if $a_1\times\s_2\in\l$ $\Rightarrow
a_1\in\l_1$, $\s_1\times a_2\in\l$ $\Rightarrow a_2\in\l_2$,
\hfill\break (P4) if $\exists\ \emptyset\ne W_i\subset {\rm
Aut}(\l_i)$;\hspace{0.1cm} $u_1\times u_2(a)\in\l,\ \forall
a\in\l,\ u_i\in W_i$.
\end{definition}

We now briefly argue why these conditions are necessary (for more
details see [7]): (P1) Since $S_1$ and $S_2$ are independent, the
state of $S$ is a product state. (P2) Let $\alpha_1\in Q_{S_1}$,
then $\alpha_1\in Q_S$ and since $S_1$ and $S_2$ are independent,
$\mu(\alpha_1)=\mu_1(\alpha_1)\times\s_2$. Moreover, let
$\alpha_2\in Q_{S_2}$. Perform $\alpha_1$ then $\alpha_2$ or
$\alpha_2$ then $\alpha_1$ or both simultaneously. Denote this
experiment as $E$. It has four possible outcomes: $yy$, $yn$, $ny$
and $nn$. Let $\alpha_1\times\alpha_2$ be the yes-no experiment on
$S$ defined by: ``perform $E$ and answer ``yes'' if one gets $yy$,
$yn$ or $ny$ and ``no'' if one gets $nn$''. Then, since $S_1$ and
$S_2$ are independent, $\mu(\alpha_1)\cup\mu(\alpha_2)=
\mu(\alpha_1\times\alpha_2)$. (P3) Let $a_1\times\s_2\in\l$ and
$\alpha\in\mu^{-1}(a_1\times\s_2)$. Then the answer ``yes'' for
$\alpha$ is certain if and only if the state $p_1$ of the first
system is in $a_1$, so that $\alpha\in Q_{S_1}$ and so
$a_1\in\l_1$. (P4) Suppose that $S_1$ and $S_2$ evolve from time
$t_0$ to time $t_1$ without interacting. Then the evolution of the
total system is given by $u_1\times u_2$ where $u_i$ is the
evolution map of system i and P4 is equivalent to condition
(\ref{daniel}). Of course, in general, not any automorphism of
$\l_i$ represents a possible evolution of system i. But any
automorphism of $\l_i$ can be interpreted as a passive action on
system i, and therefore P4 should hold for any automorphism of
$\l_i$. Moreover, if $\l_i=\pro(\h_i)$ the lattice of closed
subspaces of a Hilbert space, we must restrict to unitary maps.
Finally, remark that if $u_i\in W_i \Rightarrow u_i^{-1}\in W_i$,
then
$$(P4) \Rightarrow u_1\times u_2\ \mbox{is an isomorphism of}\ \l,\
\forall u_i\in W_i\ .$$

Assume now that $\l_1$, $\l_2$ and $\l$ are cao-lattices. Note
$\s_i$ ($\s$) the atom space of $\l_i$ (of $\l$) and $\p_i$ ($\p$)
the orthogonality relation on $\s_i$ (on $\s$). Then we assume
that \hfill\break (P5)  $p_1\p_1 q_1$ or $p_2\p_2 q_2\Rightarrow
(p_1,p_2)\p(q_1,q_2)$.

\noindent This assumption comes from relation (\ref{orthaerts}):
let $p_1,\ q_1\in\s_1$ be two orthogonal states of $S_1$, that is
there exists $\alpha\in Q_{S_1}$ such that $p_1\in\mu(\alpha)$ and
$q_1\in\mu(\alpha^\sim)$. Let $r_2,\ s_2\in\s_2$ be two arbitrary
states of $S_2$. Since $\alpha$ is a question on $S$ and $S_1$ and
$S_2$ are independent, from (\ref{orthaerts}) we ask that
$(p_1,r_2)\p(q_1,s_2)$ for all $r_2,\ s_2\in\s_2$ and $p_1\p_1
q_1\in\s_1$. But again, it is delicate to give physical arguments
for this relation to induce an ortho-complementation on $\l_{ind}$
(see [7]).\newpage

\begin{remark}\rm
Conditions P1 to P5 can easily be generalized for $n$ independent
quantum systems. P2 then reads $a_1\times\s_2\times\s_3\cdots \cup
\s_1\times a_2\times \s_2\cdots \in\l$, P3: $a_1\times\s_2\times\s_3
\cdots\in\l\ \Rightarrow a_1\in\l_1\ \cdots$ and P5:
($\exists\ j$ with $p_j\p_j q_j)
\Rightarrow (p_1,\cdots,p_n)\p(q_1,\cdots,q_n)$.
\end{remark}


\section{Results}

In the eighties, D. Aerts proposed a model for $\l_{ind}$, called
the separated product [1]. His approach was to give explicitly,
from $Q_1$ and $Q_2$, the set $Q$ of all possible yes-no
experiments on the total system. The separated product is defined
as follows:

\begin{definition}\label{depri}\rm
Let $\l_1$ and $\l_2$ be cao-lattices.\hfill\break (1) Let
$p,q\in\s_1\times\s_2$, $p\#q\Leftrightarrow p_1\p q_1$ or $p_2\p
q_2$,\hfill\break (2) $\l_1\a\l_2:=\{a\subset\s_1\times\s_2;\
a^{\#\#}=a\}$ .
\end{definition}

\begin{remark}\rm First $\l_1\a\l_2$ is a cao-lattice [6]. Second,
let $p\in\s_1\times\s_2$, then $p^\#=p_1^{\p_1}\times\s_2\cup
\s_1\times p_2^{\p_2}$.\end{remark}

In section \ref{proofs}, we will prove the following results (we
say that $W_i\subset {\rm Aut}(\l_i)$ is transitive if $\forall
p,q\in\s_i$, $\exists\ u_i\in W_i$; $u_i(p)=q$, and $\h_i$ are
Hilbert spaces over the complex numbers).
\begin{theorem}\label{nohilbert}Let $\l_i=\pro(\h_i)$ (with
${\rm dim}(\h_i)>1$) and $\l$ be a cao-lattice. Let $U(\h_i)$ the
group of unitary maps on $\h_i$. Then, $\l$ is P1, P2, P3 and P4
with $W_i=U(\h_i)$ $\Rightarrow$ $\l$ does not have the covering
property and $\l$ is not ortho-modular.
\end{theorem}
\begin{theorem}\label{thetheorem} Let $\l_1$, $\l_2$ and $\l$ be
cao-lattices. Suppose that ${\rm Aut}(\l_i)$ is transitive. Then,
$\l$ is P1, P2, P3, P4 with $W_i$ transitive and P5
$\Leftrightarrow$ $\l=\l_1\a\l_2$.
\end{theorem}
\begin{theorem}\label{plus} Let  $\l_i=\pro(\h_i)$ and $\l$ be a
cao-lattice. Then, $\l$ is P1, P2, P3 and (P4*) $u_1\times u_2$ is
an ortho-isomorphism of $\l$, $\forall\ u_i\in U(\h_i)$
$\Leftrightarrow$ $\l=\l_1\a\l_2$.
\end{theorem}

Finally, in section \ref{examples} we prove that for
$\l_i=\pro(\h_i)$, axioms P2, P3, P4 with $W_i=U(\h_i)$ and P5 are
independent. \vspace{0.2cm}

Theorem \ref{nohilbert} asserts that no Hilbert space description is possible
for two independent quantum systems. Aerts proved a similar result
for the
separated product [1] (see also [6]) and more generally for independent
systems in [2]
(see also [7]).

Assumption P4* may appear natural for $u_i$ ortho-isomorphisms,
but, again, its physical justification is delicate: if the
ortho-complementation of $\l_{ind}$ is induced by
(\ref{orthaerts}), and if two final states are orthogonal at time
$t_1$ then the yes-no experiment: ``evolve the system from time
$t_0$ to time $t_1$ and perform $\alpha$'' makes the two initial
states at time $t_0$ orthogonal.

The separated product has been investigated in [6]. It is proved
that $\l_1\a\l_2$ is irreducible $\ \Leftrightarrow\ $ $\l_1$ and
$\l_2$ are irreducible. Moreover, if $\l_1$ and $\l_2$ have the
covering property, atomic endomorphisms (join-preserving maps
sending atoms to atoms, that is evolution maps) preserve
irreducible components and factor through the components: let $\l$
be an irreducible cao-lattice having the covering
property\break\newpage \noindent and let $f$ be an atomic
endomorphism of $\l\a\l$ with the image not contained in $\l$.
Then there exist two atomic endomorphisms $f^i$ of $\l$ and a
permutation $\sigma$ such that $f=\sigma(f^1\times f^2)$ on the
atoms.

\section{Discussion and further questions}\label{discussion}

Consider two quantum systems described by $\h_1=\h_2=\C^2$
(two q-bits).
In ordinary quantum mechanics, the evolution is
given by a unitary map $u$ on $\C^2\otimes\C^2$.
If the two systems are initially independent, one always assumes the
restriction $$u: \s_1\times\s_2\rightarrow (\h_1\otimes\h_2)^*/\C$$
to be the evolution map from time $t_0$ to time $t_1$ ($\s_i=\C^{2*}/\C$).
This assumption together with condition (\ref{daniel}) imposes that
$u^{-1}(V)\cap\s_1\times\s_2\in\l_{t_0}$ for any closed subspace $V$
of $\C^2\otimes\C^2$, that is $\l_0:=\{V\cap\s_1\times\s_2;\
V\subset (\C^2\otimes\C^2)^*/\C, V^{\p_\otimes\p_\otimes}=V\}\subset\l_{t_0}$,
where $\p_\otimes$ is the orthogonality relation in the tensor product.

In proposition \ref{l0} we prove that $\l_0\supset\l_1\a\l_2$  but
$\l_0\ne\l_1\a\l_2$ (where $\l_i=
\pro(\C^2)$). Moreover, $\l_0$ has no
ortho-complementation. As a consequence, if the above description
of the interacting q-bits that are initially independent is imbued with
physical reality, the property lattice of independent quantum systems
$\l_{ind}$
is not the separated product. Moreover, in [6] we have proved that
in the separated product, no model is possible for
two interacting quantum systems that are independent before and after the
interaction takes place. Remark that this shortcoming should for
most physicists
surprisingly not be an argument to exclude the separated product as a
candidate for $\l_{ind}$ since ordinary two-body quantum theory excludes
this situation.

Nevertheless, the separated product do not allow any interaction
for quantum systems that are initially independent and therefore
can be excluded. Thus, as a consequence of theorem
\ref{thetheorem}, the assumption that relation (\ref{orthaerts})
induces an ortho-complementation on $\l$ is wrong for independent
quantum systems. The question that follows naturally is whether it
is possible for independent quantum systems to assume that $\l$ is
ortho-complemented. In  proposition \ref{riviere}, we give an
example of a cao-lattice $\l^5$ that satisfies properties P1 to P4
but not P5. But as a p-lattice, $\l^5$ is equal to the separated
product. As a consequence, we propose the following question:

\begin{question}\label{thequestion}\rm Let $\l_1=\l_2=\pro(\C^2)$. Does
there exist a cao-lattice $\l$ that is P1, P2, P3 and P4 with
$W_1=W_2=U(\C^2)$ and as a p-lattice
$\l\ne\l_1\a\l_2$?\end{question}
\section{Proofs}\label{proofs}

\demo{of Theorem \ref{nohilbert}} (1) Let $\l$ be a cao-lattice.
Let $\z(\l)$ denote the center of $\l$, and for an atom $p$, let
$\z(p)$ be the central cover of $p$. Let $p\ne q$ be two atoms.
Recall that if $\l$ has the covering property, then
$\z(p)=\z(q)\Leftrightarrow p\vee q\ne \{p,q\}$ (see [6] or
[8]).\newpage

Let $p,q\in\s_1\times\s_2$ with $p_1\ne q_1$ and $p_2\ne q_2$.
Then by P2 (we drop the subscripts $\ _1$ and $\ _2$ when no
confusion can occur), \be\ba{lll}p\vee q&=&\lcr p_1\times
\s\cap\s\times p_2\rcr\vee\lcr q_1\times \s\cap\s\times q_2\rcr\es
&\subset& \lcr p_1\times \s\vee\s\times q_2\rcr\cap \lcr q_1\times
\s\vee\s\times p_2\rcr\es &=& \lcr p_1\times \s\cup\s\times
q_2\rcr\cap \lcr q_1\times \s\cup\s\times p_2\rcr\es &=&\{p,q\}\
.\ea\label{rod}\ee

Suppose that $\l$ has the covering property. Let $p$ be an atom of
$\l$. By (\ref{rod}), we can assume that $\z(p)=r\times b$, where
$r$ is an atom and $b\subset \s_2$. But by P4, for any $u\in
W_1\times W_2$, $u(\z(p))\in\z(\l)$, and therefore, since $W_i$
are transitive, $r\times\s\in\z(\l), \forall r\in\s_1$. Thus, by
P3, $\l_1=2^{\s_1}$, which is a contradiction.

(2) Suppose that $\l$ is ortho-modular. Then $p\vee q=\{p, q\}$
$\Rightarrow\ p\p q$ or $p=q$ (see [1] or [6]). Thus by
(\ref{rod}), $p^\p\supset p_1^c\times p_2^c$ (where
$p_1^c=\s_1\backslash p_1$), $\forall p\in\s$. Let $u\in W_1\times
W_2$, then by P4 $u(p^\p)$ is a coatom and so $u(p^\p)\supset
u_1(p_1)^c\times u_2(p_2)^c\cup g_u(p)_1^c\times g_u(p)_2^c$ where
$g_u(p)=u(p^\p)^\p$. As a consequence, $u(p^\p)=u(p)^\p$ $\forall
u\in W_1\times W_2$ and $\forall p\in\s$ (see the proof of theorem
\ref{thetheorem}, part 1) and by theorem \ref{plus},
$\l=\l_1\a\l_2$. Finally, it is known that if $\l_1\a\l_2$ is
ortho-modular, then $\l_i=2^{\s_i}$ for $i=1$ or 2 (see [1] or
[6]). \cqfd

\demo{of Theorem \ref{thetheorem}} \underbar{$\Leftarrow$}: Let
$a_i\subset\s_i$. Then, by Definition \ref{depri},
$$(a_1\times \s)^{\#\#}=(a_1^\p\times \s)^\#=a_1^{\p\p}\times
\s$$ and $$ [a_1\times \s\cup\s\times a_2]^{\#\#}=(a_1^\p\times
a_2^\p)^\#= a_1^{\p\p}\times \s \cup \s\times a_2^{\p\p}\ .$$
Finally, put $W_i={\rm Aut}(\l_i)$. Let $u_1 \times u_2\in
W_1\times W_2$ and $p\in\s$. Then
$$ u_1\times
u_2(p^\#)=u_1(p_1^\p)\times \s\cup \s\times u_2(p_2^\p)\in\l_1\a\l_2\
.$$

\noindent\underbar{$\Rightarrow$}: (0) Remark that P2
$\Rightarrow\ p^\#\in\l,\ \forall p\in\s$. (1) Let $p\in\s$,
define $p_\#:=\{q\in\s;\ q^\#\subset p^\p\}$. Then by P3 and P4,
$\vert p_\#\vert\le 1$, $\forall p\in\s$. Indeed, suppose that
$\vert p_\#\vert>1$. Let $p_{\#1}:=\{q_1\in\s_1; \ q_1\times
\s\subset p^\p\}$ and, for $r\in\s_1$,
$$C_r(p^\p):=r\times \s\cap p^\p\backslash p_{\#1}\times \s\ .$$
So
$$p^\p=p_{\#1}\times \s\bigcup_{r\in\s_1\backslash p_{\#1}}
C_r(p^\p)\ .$$\newpage Suppose for instance that
$p_{\#1}\not\in\l_1$. Let $u_0\in W_1$. We have
$$\ba{l}
\bigcap\{u_0\times u(p^\p);\ u\in W_2\}\es
\displaystyle\hspace{0.0cm}=\bigcap\lac u_0(p_{\#1})\times \s\cup
\bigcup_{r\in\s_1\backslash
p_{\#1}}u_0(r)\times u(C_r(p^\p)_2);  u\in W_2\rac\es
\displaystyle\hspace{0.0cm}=u_0(p_{\#1})\times \s
\bigcup_{r\in\s_1\backslash
p_{\#1}} u_0(r)\times \bigcap_{u\in
W_2}u(C_r(p^\p)_2).\ea$$

\noindent By definition, for any $r\in\s_1\backslash p_{\#1}$,
$\exists s\in\s_2;\ C_r(p^\p)_2\subset s^c:= \s_2\backslash s$. As
a consequence, since by assumption $W_2$ is transitive,
$$\bigcap_{u\in
W_2}u(C_r(p^\p)_2)\subset \bigcap_{u\in W_2}u(s)^c= \emptyset\ .$$
\noindent By P4, we have that $u_0(p_{\#1})\times\s\in\l$, and by
P3, $u_0(p_{\#1})\in\l_1$, which is a contradiction since by
assumption $p_{\#1}\not\in\l_1$. Thus we have proved that \be
p^{\#\p}\cap q^{\#\p}= \emptyset,\ \forall p\ne q\
.\label{rodh}\ee

(2)  By P5 $p^\#\subset p^\p$,
$\forall p\in\s$, that is $p\in p^{\#\p}$. As a consequence,
(\ref{rodh}) implies that
$p^{\#\p}=\{ p\}$, $\forall p\in\s$.\cqfd

\demo{of Theorem \ref{plus}} Denote $U(\h_i)$ by $W_i$. First,
since $W_i$ are transitive, P4*$\Rightarrow\cup\{r^{\#\p};\
r\in\s\}=\s$. Let $p\in q^{\#\p}$ and let $G_p:=\{ u\in W_1\times
W_2;\ u(p)=p\}$. Then, by P4*, $p\in u(q)^{\#\p}$, $\forall u\in
G_p$, thus by (\ref{rodh}), $u(q)=q$, $\forall u\in G_p$, that is,
if ${\rm dim}(\h_i)\geq 3$ for $i=1$ and 2, $q=p$ and by
(\ref{rodh}), $p^{\#\p}=\{p \}$, $\forall p\in\s$. The case where
$\h_1=\C^2$ or $\h_2=\C^2$ is a simple extension. \cqfd


\section{Examples with $\l_1=\l_2=\pro(\C^2)$}\label{examples}

Let $\l_1=\l_2=\pro(\C^2)$ the lattice of subspaces of $\C^2$ and
$U(\C^2)$ the group of unitary maps on $\C^2$. Then
$\s_i={\C^2}^*/\C$.  We give four examples $\l^2,\cdots \l^5$ of
cao-lattices such that $\l^j$ satisfies properties P1 to P4 (with
$W_i=W_i^j$) and P5  but not property $\rm{P}j$, where
$W_i^2=W_i^3=U(\C^2)$, $W_i^5={\rm Aut}(\pro(\C^2))$ and $W_i^4$
is transitive. Finally, we give an example of a p-lattice $\l_0$
that is not ortho-complemented and satisfies properties P1, P2, P3
and P4 with $W_i={\rm Aut}(\pro(\C^2))$ and $\l_0\ne\l_1\a\l_2$.

$\l^j:=\{a\subset \s=\s_1\times\s_2;\ a^{\p_j\p_j}=a\}$ where
$\p_j$ is an orthogonality relation, that is an anti-reflexive
symmetric separating (\cal{i.e.} $p^{\p_j\p_j}=p$, $\forall p\in\s$)
binary relation on $\s$.

For $q\in\s_i$, we denote $C(q):=\{r\in\s_i;\ \vert\langle Q,
R\rangle\vert=\sqrt{3}/2\}$ where $Q\in q$, $R\in r$ and $\vert
Q\vert=\vert R\vert =1$. Finally, remark that for $u\in U(\C^2)$,
$u(C(q))=C(u(q))$.\newpage

\begin{lemma}\label{pres}
Let $\l_1$, $\l_2$ be cao-lattices and $\l$ a p-lattice.
\hfill\break (1) Suppose that $\l$ is P1. Then $\l$ is P2
$\Leftrightarrow p^\#\in\l,\ \forall p\in\s$.\hfill\break (2) Let
$\p$ be an anti-reflexive symmetric  binary relation on
$\s=\s_1\times\s_2$. If $p^{\#\p\p}\subset p^\#$, $\forall
p\in\s$, then $\p$ is an orthogonality relation.
\end{lemma}

\demo{} (1) $\Rightarrow$ : follows from P2 by definition.
$\Leftarrow$ : If $p^\#\in\l$, $\forall p\in\s$, then
$\l_1\a\l_2\subset \l$, so by Theorem \ref{thetheorem}, $\l$ is
P2.

(2) Let $p\in\s$, then $$p^{\p\p}=
(\cap\{q^{\#\p\p};\ p\in q^\#\})^{\p\p}=
\cap\{q^{\#};\ p\in q^\#\}=p\ ,$$ because for any $a\subset \s$,
$a^{\p\p\p}=a^\p$.\cqfd

\begin{lemma}\label{arch} Let $\l_1$, $\l_2$ and $\l$ be cao-lattices.
Suppose that $\l$ is P1. If $p_{\#i}\in\l_i$, $\forall p\in\s$ and
$i=1,2$, then $\l$ is P3 (where $p_{\#i}$ is defined in the proof
of Theorem \ref{thetheorem}).
\end{lemma}

\demo{} Since by assumption $p_{\#1}\in\l_1$, $\forall p\in\s$, if
$a\times\s=\cap\{p^\p;\  a\times\s\subset p^\p\}$, then
$a\in\l_1$.\cqfd

\begin{proposition}\label{these}
For $p\in\s=\s_1\times\s_2$,
define $p^{\p_2}:=p^\#\cup C(p_1)\times C(p_2)$. Then $\l^2$ is a cao-lattice
and $\l^2$ is P3, P4 with $W_i=U(\C^2)$ and P5 but not P2.
\end{proposition}

\demo{} (1) We check that $\p_2$ is an orthogonality relation: (i)
By definition, $\p_2$ is anti-reflexive.
(ii) Since $q\in C(p)\Rightarrow p\in C(q)$, $\p_2$ is symmetric.
(iii) Finally
\be(p_1^\p\times\s)^{\p_2}=\displaystyle
\bigcap_{q\in\s_2}(p_1^\p,q)^{\p_2}\subset
\bigcap_{q\in\s_2}p_1\times\s\cup \s\times q^c=p_1\times\s\ ,\label{mcgill}\ee
thus $p^{\p_2\p_2}=p$, $\forall p\in\s$.

(2) By definition, $\l^2$ is P5 and P4 with $W_i=U(\C^2)$,
since $u(p^{\p_2})=u(p)^\#\cup C(u_1(p_1))\times C(u_2(p_2))=u(p)^{\p_2}$.
By
lemma \ref{arch}, $\l^2$ is P3.

(3) Finally, $\l^2$ is not P2 because, $p^\#\not\in\l^2$. Indeed,
by (\ref{mcgill}), $p^{\#\p_2}=p$ so that
$p^{\#\p_2\p_2}=p^{\p_2}\ne p^\#$. Remark that $\l^2$ is not P2
also as a consequence of Theorem \ref{thetheorem} or Theorem
\ref{plus}.\cqfd

\begin{proposition}
For $p\in\s=\s_1\times\s_2$,
define $p^{\p_3}:=p^\#\cup C(p_1)\times\s_2\cup\s_1\times C(p_2)$.
Then $\l^3$ is a cao-lattice and $\l^3$ is P2, P4 with $W_i=U(\C^2)$ and P5
but not P3.
\end{proposition}

\demo{} (1) For the same reasons as in Proposition \ref{these},
$\p_3$ is anti-reflexive and symmetric and $\l^3$ is P4 with
$W_i=U(\C^2)$. By part (2) and Lemma \ref{pres},
 $\p_3$ is an orthogonality relation. By definition,  $\l^3$ is P5.

(2) $\l^3$ is P2: Indeed, let $\Omega=C(p_1^\p)\times C(p_2^\p)$, then
$$\displaystyle\ba{lll}p^{\#\p_3\p_3}&\subset&\displaystyle\bigcap\{q^{\p_3}=
(q_1^\p\cup C(q_1))\times\s\cup\s\times(q_2^\p\cup C(q_2));\ q\in
\Omega\}\es&=& \displaystyle\bigcup_{\omega\subset\Omega}
\bigcap_{q\in\omega}\{q_1^\p\cup C(q_1)\}\times
\bigcap_{r\in\Omega\backslash\omega}\{r_2^\p\cup C(r_2)\}= p^\#\
,\ea$$ since if $q\ne r\ne s\ne q\in C(p_i^\p)$, then $C(q)\cap
C(r)\cap C(s)=\{p_i^\p\}$.\newpage

(3) By definition, $\vert p_\#\vert >1$, thus $\l^3$ is not P3
(see the proof of Theorem \ref{thetheorem}, part 1). Remark that
$\l^3$ is not P3 also as a consequence of Theorem \ref{thetheorem}
or Theorem \ref{plus}.\cqfd

\begin{proposition}\label{john} Let $A_i\subset \s_1$ and
$g_i:A_i\rightarrow\s=\s_1\times\s_2$ be bijections ($i=1,2,3,4$)
such that $A_i\cap A_j=\emptyset$ and $A_1\cup\cdots \cup A_4=\s_1$.
Suppose moreover that $\forall i$,
\be g_i(p)\cap (p,p)^\#=\emptyset\ \forall p\in A_i\label{cond1}\ .\ee
Define $f:\s\rightarrow\s\cup\{\s\}$
by $f(p)=\s$ if $p_1\ne p_2$ and $f(p,p)=g_i(p)$ if $p\in A_i$.
Put $$p^{\p_4}:=p^\#\cup f(p)^\#\cup f^{-1}(p^\#)\ .$$
Then, $\l^4$ is a cao-lattice, $\l^4$ is P2, P3 and P5 but not P4 for any
transitive $W_i$.
\end{proposition}

\demo{} We want $\l^4$ to be P2, P3 and P5 but different from
$\l_1\a\l_2$ so that by Theorem \ref{thetheorem}, $\l^4$ is not
P4. By P5, $p^{\p_4}=p^\#\cup a_p$ where $a_p\subset \s$. So
$a_p\ne\emptyset$ at least for one $p$. Since we want $\l^4$ to be
P2, there must be at least one $r\ne p$ with $p^\#\subset
r^{\p_4}$. Further, since $\p_4$ must be symmetric, $r\in
(p_1^\p,y)^{\p_4},\ (x,p_2^\p)^{\p_4}$ for any $x$ and $y$. In
this example we choose to add one additional coatom of the
separated product only to the ortho-complement of symmetric atoms
({\it i.e.} of the form $(p,p)$).

(1) By assumption (\ref{cond1}), $\p_4$ is anti-reflexive. Further, let
$q\in f(p)^\#$, then $p\in f^{-1}(q^\#)$ and if $q\in f^{-1}(p^\#)$ then
$p\in f(q)^\#$. Thus $\p_4$ is symmetric.

(2) $\l^4$ is P2: $p^\#\subset q^{\p_4}\Leftrightarrow q=p$ or
$q\in f^{-1}(p)$. Let $\omega\subset f^{-1}(p)$ (note that $\vert
f^{-1}(p)\vert=4$). Since the atoms in $\omega$ are symmetric, if
$w$ has more than two elements, then
$$\bigcap_{q\in\omega}q^\#=\emptyset\ \mbox{thus}\
\bigcap_{q\in\omega}f^{-1}(q^\#)=f^{-1}(\cap_{q\in\omega}q^\#)=
\emptyset\ .$$ Moreover, since the inverse image by $f$ of an atom
contains only symmetric atoms,
$$\bigcap_{q\in\omega}f^{-1}(q^\#)\bigcap_{q\in f^{-1}(p)
\backslash\omega}q^\#=
\emptyset$$ if $\omega$ has two elements. As a consequence,
$$\bigcap_{q\in f^{-1}(p)}q^{\p_4}=p^\#,\
\forall p\in\s\ .$$ Thus, by Lemma \ref{pres}, $\l^4$ is a
cao-lattice and $\l^4$ is P2.

(3) $\l^4$ is P3: let $r^\p\ne s^\p\in\s_1$. Then
$\{r^\p,s^\p\}\times\s\subset q^{\p_4}\Leftrightarrow (q=(r,r)$ and
$f(r)_1=s)$ or $(q=(s,s)$ and $f(s)_1=r)$. But
$$f^{-1}((r,r)^\#)\cap f^{-1}((s,s)^\#)\supset f^{-1}(r^\p,s^\p)\ne\emptyset$$
thus $\{r^\p,s^\p\}\times\s\not\in\l^4$.\newpage

(4) By definition, $\l^4$ is P5 and $\l^4\ne\l_1\a\l_2$, so that
by Theorem \ref{thetheorem}, $\l^4$ is not P4 for any transitive
$W_i$. \cqfd

\begin{example}\rm
Let $h:\s_1\rightarrow\s_1$ be defined by $h(p)=p^\p$. Let $E_1,\
\cdots E_4\subset\s_1$ (non countable) with $E_i\cap
E_j=\emptyset$ if $i\ne j$, $\cup_{i=1}^4E_i=\s_1$ and
$h(E_j)=E_{(j+2)}$ where $(.)=.\ -1\  {\rm mod} 4 +1$.

Let $A_i^k\subset\s_1$ ($i,k=1\cdots 4$) with $A_i^k\cap A_j^k=\emptyset$ if
$i\ne j$ and $\cup_{i=1}^4A_i^k=E_k$. Moreover let
$$g_i^k:A_i^k\rightarrow E_k\times E_k\cup E_k\times E_{(k+1)}\cup
E_{(k+1)}\times E_k\cup E_{(k+1)}\times E_{(k+3)}$$
be bijections.

Define $A_i=\cup_{k=1}^4A_i^k$ and $g_i:A_i\rightarrow \s$ by
$g_i(p)=g_i^k(p)$ if $p\in A_i^k$. Then $\{(A_i,g_i)\}_{i=1\cdots
4}$ satisfies all conditions of Proposition \ref{john}.
\end{example}

\begin{proposition}\label{riviere} Let $f$ be a bijection of $\s_1$
with\hfill\break(i) $f\ne id$ \hfill\break(ii)
$f^{-1}(p^\p)=f(p)^\p$ \hfill\break(iii) $f(p)\ne p^\p$, $\forall
p\in\s_1$. Define $p^{\p_5}=(f\times f)(p)^\#$ $\forall p\in\s$.
Then $\l^5$ is a cao-lattice and $\l^5$ is P2, P3 and P4 with
$W_i={\rm Aut}(\pro(\C^2))$, but $\l^5$ is not
P5.\end{proposition}

\demo{} (1) By assumption (iii), $\p_5$ is anti-reflexive and by
assumption (ii), $\p_5$ is symmetric. Moreover, since $f\times f$
is bijective, by Lemma \ref{pres}, $\l^5$ is a cao-lattice and
$\l^5$ is P2. As a p-lattice, $\l^5=\l_1\a\l_2$, so that $\l^5$ is
P3 and P4 with $W_i={\rm Aut}(\pro(\C^2))$.

(2) By assumption (i), $\l^5$ is not P5.\cqfd

\begin{example}\rm
For $q\in\s_1$, write $q=\C(r,c(r)\e^{\i\theta})$, with $r\in
[0,1]$ and $c(r)=\sqrt{1-r^2}$. Define for $r\ne 0,1$ and
$\theta\in[0,\pi[$, $f(q):=\C(r^2,c(r^2)\e^{\i\theta})$, for
$\theta\in[\pi,2\pi[$, $f(q):=\C(\sqrt{1-c(r)},$\
$(1-r^2)^{1/4}\e^{\i\theta})$, $f(\C(1,0)):=\C(1,0)$ and
$f(\C(0,1)):=\C(0,1)$. Then $f$ is a bijection that satisfies
conditions (i) to (iii).
\end{example}

\begin{proposition}\label{l0} Let $\l_0:=\{V\cap\s=\s_1\times\s_2;\
V\subset (\C^2\otimes\C^2)^*/\C, V^{\p_\otimes\p_\otimes}=V\}$
where $\p_\otimes$ is the orthogonality relation in the tensor
product. Then $\l_0$ is a p-lattice, $\l_0$ is P1, P2, P3 and P4
with $W_i={\rm Aut}(\pro(C^2))$ and $\l_0$ is not
ortho-complemented.
\end{proposition}

\demo{} (1) $a\in \l_0\Leftrightarrow a\in \l_1\a\l_2$ or
$a=\{p,q,r\}$ with $p,q,r\in \s$ and $p_i\ne q_i\ne r_i\ne p_i$
for $i=1,2$. First, $a\in \l_1\a\l_2\Rightarrow a=\emptyset,\ \s,\
p,\ p^\#,\ p_1\times\s,\ \s\times p_2$ or $\{p,q\}$ where
$p,q\in\s$, $p_1\ne q_1$ and $p_2\ne q_2$ [6].

Let $V\subset (\C^2\otimes\C^2)^*/\C\ \mbox{with}\
V^{\p_\otimes\p_\otimes}=V$, and $x=V\cap\s$. Let $k={\rm
dim}(\langle x\rangle)$ where $\langle x\rangle$ is the subspace
of $\C^2\otimes\C^2$ spanned by $x$ and let $a=\{p^1,\cdots p^k\in
x\}$ be a basis of $\langle x\rangle$. Remark that since
$x^{\p_\otimes}=a^{\p_\otimes}$, $x^{\#}=a^{\#}$ and so
$x^{\#\#}=a^{\#\#}$. If $k=1$, then $x\in\l_1\a\l_2$. If $k=2$,
then either $p_1^1=p_1^2$ or $p_2^1=p_2^2$ or $p^1_i\ne p^2_i$ for
$i=1,2$. As a consequence, $a^{\#\#}\subset x$ and so
$x\in\l_1\a\l_2$. Finally, if $k=3$, either $a^{\#\#}$ is a coatom
of $\l_1\a\l_2$ and so $a^{\#\#}\subset x$, or $p^1_i\ne p^2_i\ne
p^3_i\ne p^1_i$ for $i=1,2$. But then $x=\{p^1,p^2,p^3\}$.\newpage

(2) $\l_0$ is not ortho-complemented:
For any $p,q,r\in \s$ with $p_i\ne q_i\ne r_i\ne p_i$ for
$i=1,2$, $\s$
covers $\{p,q,r\}$. Suppose that $\l_0$ is ortho-complemented, note
$r^\p=\{p^1,q^1,r^1\}$
and $s^\p=\{p^2,q^2,r^2\}$ with $\{ p^1,q^1, r^1\}
\cap \{ p^2,q^2, r^2\}=\emptyset$. Then $r\vee s=\s$
what is a
contradiction (see part 1). \cqfd


\noindent{\bf Acknowledgments}

\bigskip I am grateful to John Harding from New Mexico State
University for the example of Proposition \ref{john}  and for his
kind hospitality. This work was supported by the Swiss National
Science Foundation and by the Fonds Birkigt (Geneva).


\bigskip\noindent {\small REFERENCES}
\begin{list}{}{\setlength{\leftmargin}{3mm}}
\item{\footnotesize [1] D. Aerts: {\it Found. Phys.} {\bf 12}
(1982), 1131-1170.
\\ \
 [2] D. Aerts: {\it Helv. Phys. Acta} {\bf 57} (1984),
421-428.\\ \ [3] G. Birkhoff and J. von Neumann: {\it Ann.  Math.}
{\bf 37} (1936), 823-843.\\ \ [4] W. Daniel: {\it Helv. Phys.
Acta} {\bf 62} (1989), 941-968.\\ \ [5] A. Einstein, B. Podolsky
and M. Rosen: {\it Phys. Rev.} {\bf 47} (1935), 777-780.\\ \ [6]
B. Ischi: {\it Internat. J. Theoret. Phys.} {\bf 39} (2000),
2559-2581.\\ \ [7] B. Ischi: {\it Fou\-nd. Phys. Lett.} {\bf 14}
(2001), 501-518.\\ \ [8] F. Maeda and S. Maeda: {\it Theory of
Symmetric Lattices,} Springer, Berlin 1970. }
\end{list}
\end{document}